\documentclass[notitlepage,12pt]{jedm}
\usepackage[table]{xcolor}
\usepackage{url}
\usepackage{hyperref}
\usepackage{graphicx}
\usepackage{amsmath}

\begin{document}

\title{Human-Centric eXplainable AI in Education}
\date{} 

\author{{\large Subhankar Maity}\\Department of Artificial Intelligence\\Indian Institute of Technology Kharagpur\\subhankar.ai@kgpian.iitkgp.ac.in \and {\large Aniket Deroy}\\Computer Science \& Engineering\\Indian Institute of Technology Kharagpur\\roydanik18@kgpian.iitkgp.ac.in}  

\newcommand{\authorFixedWidth}[1]{\parbox[t]{.25\textwidth}{\raggedright#1 \raisebox{0pt}[0pt][6pt]{}}}

\maketitle

\begin{abstract}
As artificial intelligence (AI) becomes more integrated into educational environments, how can we ensure that these systems are both understandable and trustworthy? The growing demand for explainability in AI systems is a critical area of focus. This paper explores Human-Centric eXplainable AI (HCXAI) in the educational landscape, emphasizing its role in enhancing learning outcomes, fostering trust among users, and ensuring transparency in AI-driven tools, particularly through the innovative use of large language models (LLMs). What challenges arise in the implementation of explainable AI in educational contexts? This paper analyzes these challenges, addressing the complexities of AI models and the diverse needs of users. It outlines comprehensive frameworks for developing HCXAI systems that prioritize user understanding and engagement, ensuring that educators and students can effectively interact with these technologies. Furthermore, what steps can educators, developers, and policymakers take to create more effective, inclusive, and ethically responsible AI solutions in education? The paper provides targeted recommendations to address this question, highlighting the necessity of prioritizing explainability. By doing so, how can we leverage AI's transformative potential to foster equitable and engaging educational experiences that support diverse learners? \\ 

{\parindent0pt
\textbf{Keywords:} Human-Centric AI, eXplainable AI (XAI), Education, Large Language Models (LLMs), Learning Outcomes, Transparency
}
\end{abstract}

\section{Introduction}

The rapid advancement of AI technologies has transformed various sectors, including education, by introducing innovative solutions that enhance teaching and learning experiences . In recent years, AI systems have increasingly been utilized for personalized learning, assessment, and feedback mechanisms \cite{e2,e3,e4}. These systems analyze student data to tailor educational content and recommendations, thereby promoting a more individualized approach to learning. Among these AI innovations, large language models (LLMs) such as GPT-3.5 \cite{e6} and BERT \cite{e7} have gained significant traction due to their impressive ability to generate human-like text, assist in question generation \cite{a6,a7,a9,a10}, and provide personalized tutoring \cite{e4}. 

The unique capabilities of LLMs extend beyond simple text generation; they can engage in complex dialogues, summarize vast amounts of information, and even adapt to the evolving needs of learners \cite{e8}. This adaptability makes them valuable tools in diverse educational contexts, from K-12 classrooms to higher education and corporate training environments. However, the complexity and opaqueness of many AI models, particularly LLMs, raise significant concerns regarding their interpretability and transparency, especially in high-stakes educational environments where decisions based on AI outputs can profoundly affect students' academic trajectories \cite{e9}. The lack of clarity surrounding how these models make decisions can lead to mistrust among educators and students alike. For example, when an LLM generates feedback or assessments, stakeholders may struggle to comprehend the underlying rationale, which could result in skepticism about the AI's recommendations \cite{e10}. This gap in understanding underscores the importance of Human-Centric eXplainable AI (HCXAI), which focuses on designing AI systems that are not only effective but also understandable and trustworthy. HCXAI prioritizes the alignment of AI technologies with the needs and values of users—students, teachers, and administrators \cite{e11}.

By emphasizing transparency and user engagement, HCXAI aims to foster an educational landscape where AI tools enhance learning without compromising the integrity of the educational process. This paper investigates the implications of HCXAI in education, particularly through the lens of LLMs, and explores the myriad benefits they offer in creating a more effective and inclusive learning environment. Furthermore, we examine the potential challenges and ethical considerations associated with the deployment of these technologies in educational settings, ultimately advocating for a balanced approach that prioritizes both innovation and user understanding.

\section{The Importance of Explainability in Education}

Explainability in AI is crucial for several reasons, each contributing to the effective integration of AI technologies in educational contexts \cite{e12}. As AI systems become more prevalent in classrooms and online learning environments, understanding how these systems operate and the rationale behind their recommendations is paramount for fostering an effective learning atmosphere \cite{e13}.

\subsection{Trust Building} Trust is a foundational element in the successful adoption of AI technologies in education \cite{e14}. Students and educators are more likely to engage with AI systems that provide clear and understandable explanations for their decisions and recommendations \cite{e14,a8}. Trust is particularly vital in educational settings, where the implications of AI-driven decisions can significantly impact student learning and motivation. For instance, when LLMs suggest personalized learning paths or generate feedback on assignments, users need to understand the reasoning behind these suggestions \cite{e15}. Without clarity, users may question the validity of the AI’s output, leading to skepticism and reluctance to adopt AI tools. By enhancing explainability, educators can foster a sense of reliability in AI systems, encouraging users to embrace AI as a supportive partner in the educational process \cite{e16}.

\subsection{Enhanced Learning Outcomes} Explainable AI can significantly contribute to improved learning outcomes by empowering educators to identify the strengths and weaknesses of their students \cite{e17}. Through the insights provided by LLMs, educators can tailor interventions that promote individual learning paths, thereby accommodating diverse learning needs \cite{e12}. For example, when LLMs analyze student responses and provide detailed feedback, they highlight specific areas where students excel or struggle. This allows educators to refine their teaching strategies and adjust their approaches to meet the needs of their students effectively. If we denote \( X \) as the set of student responses, \( f(X) \) as the function representing the LLM, and \( Y \) as the personalized feedback, then:
\[
Y = f(X)
\]

This equation signifies the transformative potential of explainable AI, as it emphasizes the relationship between student input and tailored feedback, fostering a more personalized educational experience. The result is an enriched learning environment where students receive the support they need to thrive academically, ultimately leading to higher engagement and better educational outcomes.

\subsection{Accountability and Ethical Considerations}

In educational settings, the decisions made by AI systems can have significant and far-reaching consequences for students, educators, and institutions. Therefore, it is essential to have systems that are accountable and transparent, allowing stakeholders to understand how and why decisions are made \cite{e18}. The ethical implications of AI in education cannot be understated; educators must be able to justify AI-generated recommendations and actions, especially when they influence grading, tracking student progress, or suggesting resources \cite{e19}. Without transparency, there is a risk of reinforcing biases or making uninformed decisions that could adversely affect student outcomes. Moreover, accountability in AI also extends to the ethical considerations of data usage and privacy \cite{e20}. As AI systems rely on vast amounts of data to function effectively, stakeholders must be assured that student data is being used responsibly and ethically \cite{e21}. Implementing explainable AI frameworks allows for a more robust discussion around these ethical concerns, enabling educators and policymakers to advocate for responsible AI practices \cite{e21}. By fostering a culture of accountability and ethical awareness, we can ensure that AI technologies serve as equitable tools for enhancing educational experiences rather than sources of mistrust or misunderstanding.

\section{Ethical Considerations}

The implementation of Human-Centric eXplainable AI (HCXAI) in education brings forth a multitude of ethical considerations that must be carefully addressed to ensure the responsible use of AI technologies in learning environments \cite{e22}. As educational institutions increasingly integrate AI systems, the implications of these technologies extend beyond technical performance, necessitating a thorough examination of the ethical frameworks that govern their deployment \cite{e23}.

\subsection{Bias and Fairness}

One of the foremost ethical concerns surrounding AI systems, including large language models (LLMs), is the potential for bias and unfair treatment of certain student demographics \cite{e24}. AI systems are often trained on historical data that may reflect existing societal biases, inadvertently perpetuating these biases in their decision-making processes \cite{e25,e26}. For instance, if a model is trained on data that favors certain demographics, it may generate recommendations or assessments that are less favorable to others, impacting their learning experience and opportunities. This can lead to significant disparities in educational outcomes, reinforcing systemic inequities that educational institutions strive to eliminate. To mitigate bias and promote fairness in educational AI applications, it is crucial to develop comprehensive frameworks that actively identify and address these issues. A common measure to assess bias is the disparity in outcomes \( O_i \) for different demographic groups \( G_j \):
\[
\text{Disparity}(O_i, G_j) = \frac{\sum_{k \in G_j} O_i(k)}{|G_j|} - \frac{\sum_{l \in G_i} O_i(l)}{|G_i|}
\]

In this equation, \( O_i(k) \) is the outcome for individual \( k \) in group \( j \), and \( |G_j| \) is the size of group \( j \). By calculating this disparity, educators and policymakers can gain insights into the fairness of AI-generated outcomes and make necessary adjustments to ensure equitable treatment across diverse student populations. Developing and implementing these frameworks not only helps in identifying biases but also fosters a culture of accountability and continuous improvement in the design and application of educational AI systems.

\subsection{Privacy and Data Security}

The use of AI in education often involves the collection and analysis of sensitive student data, including personal information, academic performance, and behavioral patterns \cite{e23}. Ensuring that this data is handled ethically and securely is paramount to maintain student privacy and trust \cite{e27}. As educational institutions adopt AI technologies, they must implement robust data protection measures to safeguard this information from breaches, misuse, or unauthorized access. Transparency in data usage is essential; stakeholders should be informed about what data is being collected, how it will be utilized, and who will have access to it \cite{e28}. Furthermore, institutions should establish clear data governance policies that delineate responsibilities for data handling and outline procedures for responding to potential data breaches. By prioritizing privacy and data security, educational institutions can build trust among students and their families, encouraging wider acceptance and engagement with AI technologies.

\subsection{Informed Consent}

Informed consent is another critical ethical requirement when integrating AI systems into educational settings \cite{e21}. Students and educators must be adequately informed about how AI systems operate, including the algorithms and data that drive their functionalities \cite{e29}. This understanding is particularly important for vulnerable populations, such as minors, who may not fully grasp the implications of using AI technologies in their learning experiences \cite{e16}. Obtaining informed consent involves ensuring that users have a clear understanding of what data is collected, how it is utilized, and the potential consequences of its use \cite{e30}. Educational institutions should provide accessible information and resources that explain AI technologies in layman's terms, allowing users to make informed decisions about their participation. By prioritizing informed consent, educators can respect the autonomy of their students and empower them to engage critically with AI systems.

\subsection{Autonomy and Agency}

Finally, it is essential to recognize that AI systems should enhance, rather than replace, human decision-making in educational contexts \cite{e31}. The introduction of AI technologies must not undermine the autonomy and agency of educators and students; instead, it should empower them to take an active role in their learning processes \cite{e32}. Educators should retain the authority to make pedagogical decisions, utilizing AI-generated insights as supportive tools rather than prescriptive mandates. Students, likewise, must feel empowered to question AI-generated outputs and engage critically with the feedback provided by these systems \cite{e33}. This autonomy fosters an educational environment that values critical thinking and encourages learners to take ownership of their educational journeys \cite{e34}. By emphasizing the importance of autonomy and agency, educational institutions can create a more balanced and collaborative approach to integrating AI technologies, ensuring that human judgment remains at the forefront of the educational experience.

\section{Challenges in Implementing HCXAI}

Despite the numerous benefits associated with Human-Centric eXplainable AI (HCXAI), several significant challenges must be addressed to ensure its effective implementation in educational settings. As educational institutions strive to harness the power of AI technologies, they must navigate these complexities to create systems that are not only functional but also accessible and beneficial for all users.

\subsection{Complexity of AI Models}

One of the primary challenges in implementing HCXAI is the inherent complexity of many state-of-the-art AI models, including large language models (LLMs) \cite{e35}. These models often operate as “black-boxes,” meaning that their internal mechanisms and decision-making processes are not easily interpretable \cite{e36,e37}. This lack of transparency poses a significant barrier to understanding how and why an LLM generated a specific response or suggestion. For educators and students, this opacity can lead to frustration and skepticism regarding the reliability of AI outputs. To overcome this challenge, researchers and practitioners must develop more interpretable AI models that can provide clear and meaningful explanations of their reasoning \cite{e37}. This may involve creating visualization tools or simpler algorithms that offer insights into model behavior without sacrificing performance. Furthermore, ongoing efforts in the field of AI explainability should focus on developing standardized metrics for evaluating the interpretability of AI systems, ensuring that users can assess and compare the explainability of different models effectively.

\subsection{Diverse User Needs}

Another significant challenge in the implementation of HCXAI is the diversity of user needs within educational environments \cite{e38}. Educators and students come with varying levels of AI literacy, which influences their capacity to understand and utilize AI-generated insights effectively \cite{e39}. For example, a seasoned educator may require a more in-depth explanation of an AI system's recommendations, while a student with limited familiarity with AI may benefit from straightforward, easy-to-understand guidance \cite{e40}. This variability necessitates the development of adaptable and user-friendly explanation interfaces that cater to the unique needs of different stakeholders. Tailoring explanations to accommodate various levels of expertise can help foster a more inclusive environment where all users feel empowered to engage with AI technologies \cite{e41}. Educators should also receive training on how to interpret and communicate AI-generated insights to their students, enabling them to act as intermediaries in bridging the gap between complex AI outputs and user understanding.

\subsection{Integration with Existing Educational Practices} 

Integrating HCXAI systems into traditional educational frameworks presents another formidable challenge \cite{e42}. Many educational institutions operate within established pedagogical structures and assessment methods that may not readily accommodate the use of AI technologies. This integration process requires thoughtful consideration of how AI systems can complement and enhance existing practices rather than disrupt them \cite{e1}. For successful integration, educational stakeholders must engage in collaborative efforts to rethink and redesign pedagogical approaches that incorporate AI tools effectively \cite{e43}. This may involve piloting HCXAI systems in controlled environments, gathering feedback from users, and iteratively refining the integration process based on real-world experiences. Additionally, institutions must consider the necessary infrastructure and support systems, such as professional development for educators, to facilitate the smooth adoption of AI technologies \cite{e44}. By addressing these integration challenges proactively, educational institutions can better position themselves to leverage the full potential of HCXAI in fostering enriched learning experiences.

\section{The Role of LLMs in HCXAI}

Large language models (LLMs) play a pivotal role in Human-Centric eXplainable AI (HCXAI) in education by offering advanced capabilities that can be harnessed to enhance explainability, improve learning outcomes, and foster meaningful interactions between students and AI systems \cite{e45}. As educational institutions increasingly integrate AI technologies, the unique attributes of LLMs can significantly contribute to creating a more transparent and effective learning environment.

\subsection{Natural Language Explanations}

One of the most significant advantages of LLMs is their ability to generate human-readable explanations for their outputs \cite{a8}. These models can articulate the rationale behind their suggestions or feedback in a manner that is understandable to users, thereby providing clarity on the decision-making processes at play \cite{e46}. For instance, in contexts such as automated grading, LLMs can explain why a particular score was assigned to a student's response, detailing the criteria considered and the aspects that met or fell short of expectations \cite{a8}. This capacity for natural language explanations is particularly useful in personalized learning recommendations, where students may be directed towards specific resources or activities based on their performance and learning styles \cite{a8}. By elucidating the reasoning behind these recommendations, LLMs not only enhance student comprehension but also build trust in the AI system. When students understand the "why" behind a recommendation, they are more likely to engage with the material and take ownership of their learning journey.

\subsection{Interactive Dialogues}
In addition to providing explanations, LLMs can engage in interactive dialogues with students, fostering a dynamic and responsive learning environment \cite{e47}. Through these interactions, students can ask clarifying questions, seek additional information, and receive tailored explanations in real time. This back-and-forth conversation model transforms the traditional one-way communication of educational content into a more engaging and participatory experience. The interaction can be modeled as follows:
\[
\text{Response}(S, Q) = f_{LLM}(S, Q)
\]

In this equation, \( S \) represents the student input, \( Q \) represents the question, and \( f_{LLM} \) is the language model function. By allowing students to direct their inquiries and receive immediate feedback, LLMs empower learners to take control of their educational experiences. This level of interactivity is particularly valuable in identifying and addressing misunderstandings, as students can clarify doubts and explore topics in depth, leading to improved retention and comprehension of complex concepts.

\subsection{Content Generation}

LLMs are also capable of assisting educators in generating educational content, including quizzes, lesson plans, and summaries, while simultaneously explaining the underlying concepts in an accessible manner \cite{e48}. This capability can significantly reduce the workload for educators, allowing them to focus on higher-order teaching tasks and student engagement. Moreover, when generating content, LLMs can incorporate explanations that connect theoretical concepts to practical applications, enhancing the relevance of the material for students \cite{e49}. For instance, an LLM can create a quiz question and provide a detailed explanation of the concept being assessed, helping students understand the rationale behind the correct answers. This dual function of content generation and explanation ensures that learning resources are not only informative but also pedagogically sound.

In summary, LLMs serve as powerful tools in the HCXAI framework by providing natural language explanations, enabling interactive dialogues, and assisting in content generation. By leveraging these capabilities, educational institutions can create a more transparent and engaging learning environment that fosters trust, enhances understanding, and empowers students to take an active role in their education.

\section{Frameworks for Developing HCXAI Systems}

To create effective Human-Centric eXplainable AI (HCXAI) systems in education, particularly those leveraging large language models (LLMs), it is essential to adopt comprehensive frameworks that prioritize user needs, diverse explanation strategies, and ongoing feedback. The following frameworks provide a structured approach to developing HCXAI systems that foster transparency, engagement, and enhanced learning outcomes.

\subsection{User-Centered Design}

A foundational principle in developing HCXAI systems is to involve educators and students directly in the design process \cite{e50}. This user-centered approach ensures that AI tools are tailored to meet the specific needs, preferences, and contexts of their users \cite{e51}. Engaging users early in the design phase allows developers to gather valuable insights into the challenges and expectations of both educators and students, facilitating the creation of systems that are more relevant and effective. Conducting usability testing is crucial in this framework. Through iterative testing, developers can refine explanations and interfaces, ensuring they are not only functional but also accessible and understandable to users with varying levels of expertise. Feedback from usability testing can help identify potential barriers to comprehension, leading to design adjustments that enhance the overall user experience. Moreover, involving educators in the design process promotes buy-in and encourages the adoption of HCXAI systems within educational settings, ultimately contributing to their success.

\subsection{Multi-Faceted Explanation Strategies}

Implementing a range of explanation methods is vital to cater to the diverse learning styles and preferences of students \cite{e52}. Different individuals process information differently, and providing a variety of explanation formats—such as visualizations, textual explanations, and interactive feedback can significantly enhance understanding \cite{e53}. For instance, some students may benefit from visual representations of data that highlight relationships between concepts, while others may prefer detailed textual descriptions that explain the reasoning behind AI outputs. Furthermore, integrating interactive elements allows students to engage actively with the content, fostering deeper comprehension and retention. In addition to varied formats, contextual explanations are essential for enhancing the relevance and comprehension of AI outputs. By relating the information generated by LLMs to specific educational objectives and real-world applications, educators can help students see the connections between AI-generated insights and their learning goals \cite{e54}. This contextualization not only improves understanding but also motivates students by illustrating the practical implications of their learning experiences.

\subsection{Continuous Feedback Mechanisms}

Establishing continuous feedback mechanisms is crucial for the iterative improvement of HCXAI systems \cite{e55}. Feedback loops allow users to provide input on the explanations and functionalities of AI systems, enabling developers to make data-driven adjustments that enhance user experience and learning outcomes. Regularly soliciting feedback from educators and students helps identify areas for improvement and fosters a culture of collaboration between users and developers. Moreover, monitoring the effectiveness of HCXAI in real-world educational settings is essential for understanding how these systems impact learning. Data collected from user interactions can be analyzed to assess the effectiveness of different explanation methods and overall system performance. This ongoing evaluation provides insights that inform future iterations of the system, ensuring that it remains responsive to user needs and evolving educational contexts.

By adopting these frameworks for developing HCXAI systems, educational institutions can create AI technologies that are not only effective but also deeply aligned with the needs and values of their users \cite{e56}. Through user-centered design, multi-faceted explanation strategies, and continuous feedback mechanisms, HCXAI can contribute significantly to enhancing the educational experience and promoting better learning outcomes for all students.

\section{Recommendations for Stakeholders}

To effectively harness Human-Centric eXplainable AI (HCXAI) in education, particularly through the integration of large language models (LLMs), it is essential for various stakeholders to take proactive measures \cite{e57}. The following recommendations outline specific actions for educators, developers, and policymakers to ensure that AI technologies are implemented responsibly and effectively in educational contexts.

\subsection{For Educators}

Educators play a crucial role in shaping the future of AI in education. To maximize the benefits of HCXAI, teachers should actively engage with AI tools that prioritize explainability \cite{e58}. This involves not only using these tools in their classrooms but also becoming advocates for transparency in AI-based decision-making processes. Educators should seek to understand how these AI systems function and the rationale behind their outputs, enabling them to better guide their students in interpreting AI-generated insights. Additionally, educators should foster a culture of critical thinking around AI technologies among their students \cite{e59}. This includes teaching students how to question and evaluate AI outputs, encouraging them to think critically about the information presented to them. By promoting digital literacy and understanding of AI systems, educators can empower students to become informed consumers of AI-generated content, ultimately enhancing their learning experiences. Furthermore, educators can collaborate with developers to provide feedback on the usability and effectiveness of AI tools in real-world educational settings. Their insights can help refine these technologies to better meet the needs of both teachers and students, ensuring that HCXAI systems are genuinely beneficial.

\subsection{For Developers} 

Developers of AI systems bear the responsibility of creating explainable AI technologies that prioritize user understanding \cite{e60}. This includes incorporating diverse explanation strategies that cater to different learning styles and preferences, as well as ensuring that explanations are not only accurate but also clear and accessible to users. Developers should engage in iterative testing and feedback loops with educators and students to refine their systems continually. Moreover, developers should consider the ethical implications of their technologies. This involves actively working to identify and mitigate biases in AI algorithms and ensuring that their systems promote fairness and inclusivity. By prioritizing ethical considerations, developers can contribute to building trust in AI systems and fostering a positive perception of technology in education. In addition to focusing on user understanding, developers should also invest in robust documentation and training resources. Providing educators with comprehensive guides on how to effectively use and interpret AI systems will enhance the overall user experience and maximize the potential of HCXAI in educational environments \cite{e61}.

\subsection{For Policymakers}

Policymakers play a vital role in shaping the landscape of AI in education \cite{e62}. To promote the responsible use of AI technologies, it is essential to develop guidelines and regulations that emphasize the importance of explainability and accountability in AI systems \cite{e63}. These regulations should ensure that educational institutions adopt AI tools that are transparent and subject to scrutiny, allowing stakeholders to understand how decisions are made and what data is being used. Policymakers should also advocate for the establishment of ethical standards in AI development and deployment \cite{e62}. This includes fostering collaboration between educators, developers, and researchers to create best practices that prioritize transparency, fairness, and accountability. By encouraging partnerships between these stakeholders, policymakers can facilitate the sharing of knowledge and resources, leading to the development of more effective and responsible AI systems. Moreover, it is crucial for policymakers to allocate resources for training and professional development in AI for educators \cite{e64}. By investing in programs that enhance educators' understanding of AI technologies and their implications, policymakers can empower teachers to make informed decisions about the use of AI in their classrooms.

In summary, the effective implementation of HCXAI in education requires coordinated efforts from all stakeholders involved. Educators should engage with and advocate for explainable AI tools, developers must focus on user understanding and ethical considerations, and policymakers should establish guidelines that promote responsible AI use. Through these collective actions, the potential of HCXAI to transform education can be fully realized.

\section{Conclusion}

Human-Centric eXplainable AI (HCXAI) holds remarkable potential to revolutionize the field of education, particularly through the innovative application of large language models (LLMs). These advanced AI technologies can significantly enhance educational experiences by fostering trust among users, improving learning outcomes, and ensuring transparency in AI-driven tools and processes. As educational environments increasingly incorporate AI technologies, the need for explainable systems becomes paramount. The successful implementation of HCXAI involves addressing the inherent challenges associated with these technologies. By adopting user-centered design principles, stakeholders—including educators, developers, and policymakers—can work collaboratively to create educational tools that are not only effective but also aligned with the diverse needs of students and teachers. A focus on explainability ensures that users can comprehend and engage with AI-generated insights, thereby enhancing their learning experiences and empowering them to take control of their educational journeys. Moreover, the integration of HCXAI into educational contexts promotes inclusivity and equity. By prioritizing transparency and accountability in AI systems, stakeholders can mitigate biases that may arise from the underlying data and algorithms, thus ensuring that all students receive fair and equitable support in their learning processes. This commitment to fairness not only builds trust but also fosters a sense of belonging among learners, which is crucial for their academic success and overall well-being.

As we look to the future, it is essential to recognize that the landscape of education is continuously evolving, influenced by the rapid advancements in AI technologies. Therefore, a proactive approach to developing HCXAI systems that emphasize explainability will be critical in shaping educational practices. By doing so, we can create AI systems that truly serve the needs of learners and educators alike, paving the way for more personalized and effective learning experiences. In conclusion, the journey toward implementing Human-Centric eXplainable AI in education is an ongoing endeavor that requires the collective efforts of all stakeholders. By prioritizing trust, transparency, and user understanding, we can harness the transformative power of AI to create a brighter future for education—one where technology enhances, rather than detracts from, the learning experience. As we embrace the potential of HCXAI, we must remain vigilant in our commitment to ethical practices, continuous improvement, and the empowerment of all learners, ensuring that the benefits of AI are accessible to everyone.
    
\bibliographystyle{acmtrans}
\bibliography{ref}

\end{document}